\documentclass[twocolumn,showpacs,preprintnumbers,amsmath,amssymb,floatfix]{revtex4-2}
\usepackage{graphicx}
\usepackage{dcolumn}
\usepackage{bm}
\usepackage{latexsym}
\usepackage[hang,small,bf]{caption}

\usepackage{braket}
\usepackage{subfigure}
\usepackage{here}

\usepackage{multirow} 

\usepackage{comment}
\newcommand{\mbra}[1] { \left\{ #1 \right\}}

\def\nn{\nonumber}
\newcommand{\beqn}{\begin{eqnarray}}
\newcommand{\beq}{\begin{equation}}
\newcommand{\eeqn}{\end{eqnarray}}
\newcommand{\eeq}{\end{equation}}

\newcommand{\tr}{\mathop{\rm Tr}}

\newcommand{\rcnp}{\affiliation{Research Center for Nuclear Physics, Osaka University, Osaka 567-0047, Japan}}

\begin{document}
\title{Monopoles of the Dirac type and color confinement in QCD \\
- Study of the continuum limit  -}
\author{Tsuneo Suzuki}
\email[e-mail:]{tsuneo@rcnp.osaka-u.ac.jp}
\rcnp

\date{\today}

\begin{abstract}
Non-Abelian gauge fields having a line-singularity of the Dirac type lead us to violation of the non-Abelian Bianchi identity. The violation as an operator is equivalent to violation of Abelian-like Bianchi identities
corresponding to eight Abelian-like conserved magnetic monopole currents of the Dirac type in $SU(3)$ QCD. It is very interesting to study if these new Abelian-like monopoles are responsible for color confinement in the continuum $SU(3)$ QCD, since any reliable candidate of color magnetic monopoles is not known yet. If these new Abelian-like monopoles exist in the continuum limit, the Abelian dual Meissner effect occurs, so that the linear part of the static potential between a quark-antiquark pair is reproduced fully by those of Abelian and monopole static potentials. These phenomena are called here as perfect Abelian and monopole dominances. It is shown that the perfect Abelian dominance is reproduced fairly well, whereas the perfect monopole dominance seems to be realized for large $\beta$ when use is made of the smooth lattice configurations in the maximally Abelian (MA) gauge. Making use of a block spin transformation with respect to monopoles, the scaling behaviors of the monopole density and the effective monopole action are studied. Both monopole density and the effective monopole action which are usually a two-point function of $\beta$ and the number of times $n$ of the block spin transformation are a function of $b=na(\beta)$ alone for $n=1,2,3,4,6,8,12$.  If the scaling behavior is seen for up to larger $n$, it  shows the existence of the continuum limit, since $a(\beta)\to 0$ when $n\to\infty$ for fixed $b=na(\beta)$. Along with the previous results without any gauge fixing, these new results obtained in MA gauge suggest that the new Abelian-like monopoles play the role of color confinement in $SU(3)$ QCD.    
\end{abstract}


\maketitle
\section{Introduction}

Color confinement in quantum chromodynamics (QCD) is still an important unsolved problem. 
As a picture of color confinement, 't~Hooft~\cite{tHooft:1975pu} and Mandelstam~\cite{Mandelstam:1974pi} conjectured that the QCD vacuum is a kind of a magnetic superconducting state caused by condensation of magnetic monopoles and  an effect dual to the Meissner effect works to confine color charges. 
However to find color magnetic monopoles which
condense is not straightforward in QCD. 
If the dual Meissner effect picture is correct, it is absolutely necessary to derive  such color-magnetic monopoles from gluon dynamics of QCD. 

An interesting idea to introduce such an Abelian monopole in QCD is to project QCD to the Abelian maximal torus group by a partial (but singular) gauge fixing~\cite{tHooft:1981ht}.  In $SU(3)$ QCD, the maximal torus group is  Abelian $U(1)^2$. Then Abelian magnetic monopoles appear as a topological object at the space-time points corresponding to the singularity of the gauge-fixing matrix. Condensation of the monopoles  causes  the dual Meissner effect with respect to $U(1)^2$. Numerically, an Abelian projection in various gauges such as the maximally Abelian (MA) gauge~\cite{Kronfeld:1987ri,Kronfeld:1987vd} seems to support the conjecture~\cite{Suzuki:1992rw, Chernodub:1997ay, Shiba:1994ab, SNW:1994}.  
Although numerically interesting, the idea of Abelian projection~\cite{tHooft:1981ht} is theoretically unsatisfactory. Especially there are infinite ways of such a partial gauge-fixing and whether the 't Hooft scheme depends on gauge choice or not is not known. 

Motivated by an interesting work by Bonati et al.\cite{Bonati:2010tz} which found volation of non-Abelian Bianchi identity (VNABI) exists behind the 'tHooft Abelian monopoles, the present author found in 2014~\cite{Suzuki:2014wya} an interesting and more fundamental fact  that, when original gluon fields have a singularity where partial derivatives are not commutative, the non-Abelian Bianchi identity is broken and VNABI is just equal to the violation of Abelian-like Bianchi identities. The latter just  corresponds to the existence of Abelian-like monopoles. For more details, see also Ref.\cite{Suzuki:20220422}.

Define a covariant derivative operator $D_{\mu}=\partial_{\mu}-igA_{\mu}$. The Jacobi identities are expressed as $\epsilon_{\mu\nu\rho\sigma}[D_{\nu},[D_{\rho},D_{\sigma}]]=0$.
By direct calculations, one gets
$[D_{\rho},D_{\sigma}]=-igG_{\rho\sigma}+[\partial_{\rho},\partial_{\sigma}]$,
where the second commutator term of the partial derivative operators can not be discarded in general, since gauge fields may contain a line singularity. Actually, it is the origin of the violation of the non-Abelian Bianchi identities (VNABI) as shown in the following. The non-Abelian Bianchi identities and the Abelian-like Bianchi identities are, respectively: $D_{\nu}G^{*}_{\mu\nu}=0$ and $\partial_{\nu}f^{*}_{\mu\nu}=0$.
The relation $[D_{\nu},G_{\rho\sigma}]=D_{\nu}G_{\rho\sigma}$ and the Jacobi identities lead us to
\begin{eqnarray}
D_{\nu}G^{*}_{\mu\nu}&=&-\frac{i}{2g}\epsilon_{\mu\nu\rho\sigma}[D_{\nu},[\partial_{\rho},\partial_{\sigma}]]\nn\\
&=&\frac{1}{2}\epsilon_{\mu\nu\rho\sigma}[\partial_{\rho},\partial_{\sigma}]A_{\nu}
=\partial_{\nu}f^{*}_{\mu\nu}, \label{eq-JK}
\end{eqnarray}
where $f_{\mu\nu}$ is defined as $f_{\mu\nu}=\partial_{\mu}A_{\nu}-\partial_{\nu}A_{\mu}=(\partial_{\mu}A^a_{\nu}-\partial_{\nu}A^a_{\mu})\lambda^a/2$. Namely Eq.(\ref{eq-JK}) shows that the violation of the non-Abelian Bianchi identities, if exists,  is equivalent to that of the Abelian-like Bianchi identities.
Denote the violation of the non-Abelian Bianchi identities (VNABI) as  $J_{\mu}=D_{\nu}G^*_{\mu \nu}$ and Abelian-like monopole currents $k_{\mu}$ without any gauge-fixing as the violation of the Abelian-like Bianchi identities:
$k_{\mu}=\partial_{\nu}f^*_{\mu\nu}
=\frac{1}{2}\epsilon_{\mu\nu\rho\sigma}\partial_{\nu}f_{\rho\sigma}.$ 
Eq.(\ref{eq-JK}) shows that 
$J_{\mu}=k_{\mu}$.
The Abelian-like monopole currents satisfy an Abelian conservation rule kinematically, $\partial_\mu k_\mu^a(x)=0$\cite{Arafune:1974uy}. There can exist  
exact Abelian (but kinematical) symmetries in non-Abelian QCD.  This is an extension of the Dirac idea~\cite{Dirac:1931} of monopoles in Abelian QED to non-Abelian QCD. 

In the framework of simpler $SU(2)$ QCD, interesting numerical results were obtained. Abelian and monopole dominances as well as the Abelian dual Meissner effect are seen clearly without any additional gauge-fixing  already in 2009~\cite{Suzuki:2007jp,Suzuki:2009xy}, although at that time, no theoretical explanation was clarified with respect to Abelian-like monopoles without any gauge-fixing. They are now found to be just Abelian-like monopoles proposed in the above paper~\cite{Suzuki:2014wya}.  Also, the existence of the continuum limit of this new kind of Abelian-like monopoles was discussed with the help of the block spin renormalization group concerning the Abelian-like monopoles.  The beautiful scaling behaviors showing the existence of the continuum limit are observed with respect to the monopole density~\cite{Suzuki:2017lco} and the infrared effective monopole action~\cite{Suzuki:2017zdh}. The scaling behaviors seem also to be independent of gauges smoothing the lattice vacuum. 
\begin{table}
\caption{The ratio of the Abelian and the non-Abelian string tensions $\sigma_a/\sigma_F$ determined by applying the multilevel method in the Wilson action. The data are cited from Ref.\cite{IHS:202207}.}
 \label{T1}   
    \centering     
\begin{tabular}{l|c|c}     
\hline
Lattice size &$\beta$ &$\sigma_a/\sigma_F$ \\
\hline         
$12^4$ & 5.6& 0.87(13)\\
\hline
$16^4$ & 5.6& 1.05(9)\\
\hline
$12^4$ &5.7& 0.91(8)\\
\hline
$12^4$&5.8& 1.01(11)\\
\hline
\end{tabular}
\end{table}

\begin{table} 
     \caption{String tensions from Polyakov-loop correlations 
     in the Wilson action at $\beta=5.6$ on $24^3 \times 4$. The data are from Ref.\cite{IHS:202207}.}
    \label{T2} 
\centering            
\begin{tabular}{l|c}     
\hline    
Types of the potential & $\sigma a^2$ \\
\hline
non-Abelian   & 0.178(1)    \\
\hline 
Abelian    & 0.16(3)     \\
\hline 
monopole  & 0.17(2)     \\
\hline 
photon   &-0.0007(1) \\
\hline       
\end{tabular}
\end{table}

It is very interesting to study the new Abelian-like monopoles in $SU(3)$ QCD.  To check if the Dirac-type monopoles are a key quantity of color confinement in the continuum $SU(3)$ QCD, it is necessary to study  monopoles numerically in the framework of lattice $SU(3)$ QCD and to study then if the continuum limit exists. It is not so straightforward, however, to extend the previous $SU(2)$ studies to $SU(3)$. How to define Abelian-like link fields and monopoles without gauge-fixing is not so simple as  in $SU(2)$, since a $SU(3)$ group link field is not  expanded in terms of Lie-algebra elements defining Abelian link fields as simply done as in  $SU(2)$. There are theoretically many possible definitions which have the same naive continuum limit in $SU(3)$.  In the previous work~\cite{IHS:202207},  we found a natural definition as shown later explicitly. Using the definition, we showed as cited in Table~\ref{T1} that 
the perfect Abelian dominance exists with the help of the multilevel method\cite{Luscher:2001up,Luscher2002} but without introducing additional smoothing techniques like partial gauge fixings. Table~\ref{T2} shows that the perfect monopole dominance holds good again without any additional gauge fixing. 
In the latter, we had to evaluate huge number of correlations between non-local gauge-variant quantites in order to extract probable gauge-invariant results\cite{Elitzur:1975}.  

The dual Meissner effect around a pair of static quark and antiquark was studied. Abelian electric fields are squeezed due to solenoidal monopole currents and the penetration length for an Abelian electric field of a single color is the same as that of non-Abelian electric field. The coherence length was also measured directly through the correlation of the monopole density and the Polyakov loop pair.
The Ginzburg-Landau parameter indicates that the $SU(3)$ vacuum in the confinement phase is that of the weak type I (dual) superconductor. But these previous results in Ref\cite{IHS:202207} are all only on very small lattices and at restricted $\beta$. However they suggest that the new idea of monopoles are also important in real $SU(3)$ QCD. It is necessary to study on larger lattices at more different $\beta$ in order to show the existence of the new monopoles actually in the continuum $SU(3)$.

 Here the aim of this note is to study the scaling behaviors of the Abelian monopoles with the help of additional  technique  reducing lattice artifact monopoles as much as possible. First the most popular partial gauge fixing, the maximally Abelian gauge is adopted for the Iwasaki improved gluon action~\cite{Iwasaki:1985,IKKY:1997,Takeda:2004}
 on $48^4$ lattices for
various coupling constants between $\beta=2.3$ and $\beta=3.5$. It is studied if the Abelian dominance and the monopole dominance expected from the Abelian dual Meissner picture\cite{Koma:2003} are realized. Next introducing the block spin transformation, we measure the renormalization flows of the monopole density and the effective monopole action and study directly if the Abelian-like monopoles have the continuum limit.

\begin{table}[htb]
\begin{center}
\caption{\label{potential_parameter}
Simulation parameters.  (The coupling constant $\beta$, 
the lattice size, the gauge configuration number, the lattice spacing $a(\beta)$ from Ref.\cite{Takeda:2004}, $ns$ the number of the Abelian smearing steps, ($t_1^{a,m}$, 
$t_2^{a,m}$) the fitting ranges in lattice unit. }
\begin{tabular}{c|c|c|c|c|c}
\hline
$\beta$ &volume& $N_{\rm conf}$& $a(\beta)$~[fm]& $ns$, $(t_1^a, t_2^a)$ & $ns$, $(t_1^m, t_2^m)$\\ 
\hline
2.9 & $48^4$ & 160 & 0.1420(116)&6, (1, 10)  &2, (5, 10)\\
3.0 & $48^4$ & 160 & 0.1312(99)&4, (3, 11)  &0, (10, 17)\\
3.1 & $48^4$ & 160 & 0.1143(46)&8, (2, 15)  &2, (5, 17)\\
3.2 & $48^4$ & 160 & 0.1080(60)&6, (10, 18)  &0, (14, 24)\\  
3.3 & $48^4$ & 160 & 0.0918(65)&8, (1, 20)  &60, (15, 20)\\
3.4 & $48^4$ & 160 & 0.0855(75) &10, (4, 18)  &0, (15, 24)\\
3.5 & $48^4$ & 160 & 0.0809(130)&12, (3, 18)  &2, (4, 20)\\
\hline
\end{tabular}
\end{center}
\end{table}

\begin{figure}[htb]
\caption{Examples of the effective mass plots at $\beta=3.5$ on $48^4$ for  the Abelian (up) and the monopole (down) parts.}
\label{effmass}
  \begin{minipage}[b]{0.9\linewidth}
    \centering
 \includegraphics[width=8cm,height=6.cm]{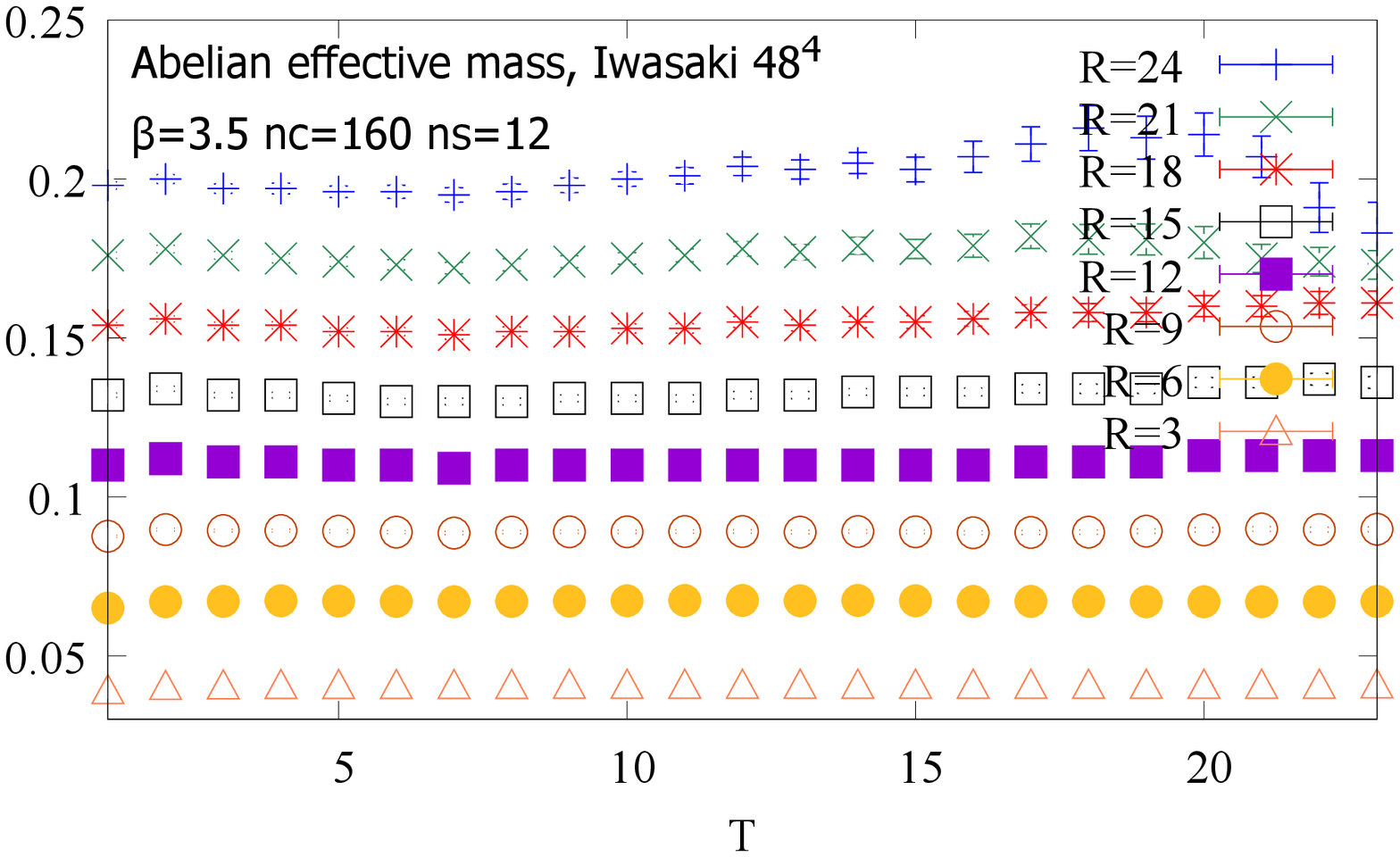}
  \end{minipage}
  \begin{minipage}[b]{0.9\linewidth}
    \centering
 \includegraphics[width=8cm,height=6.cm]{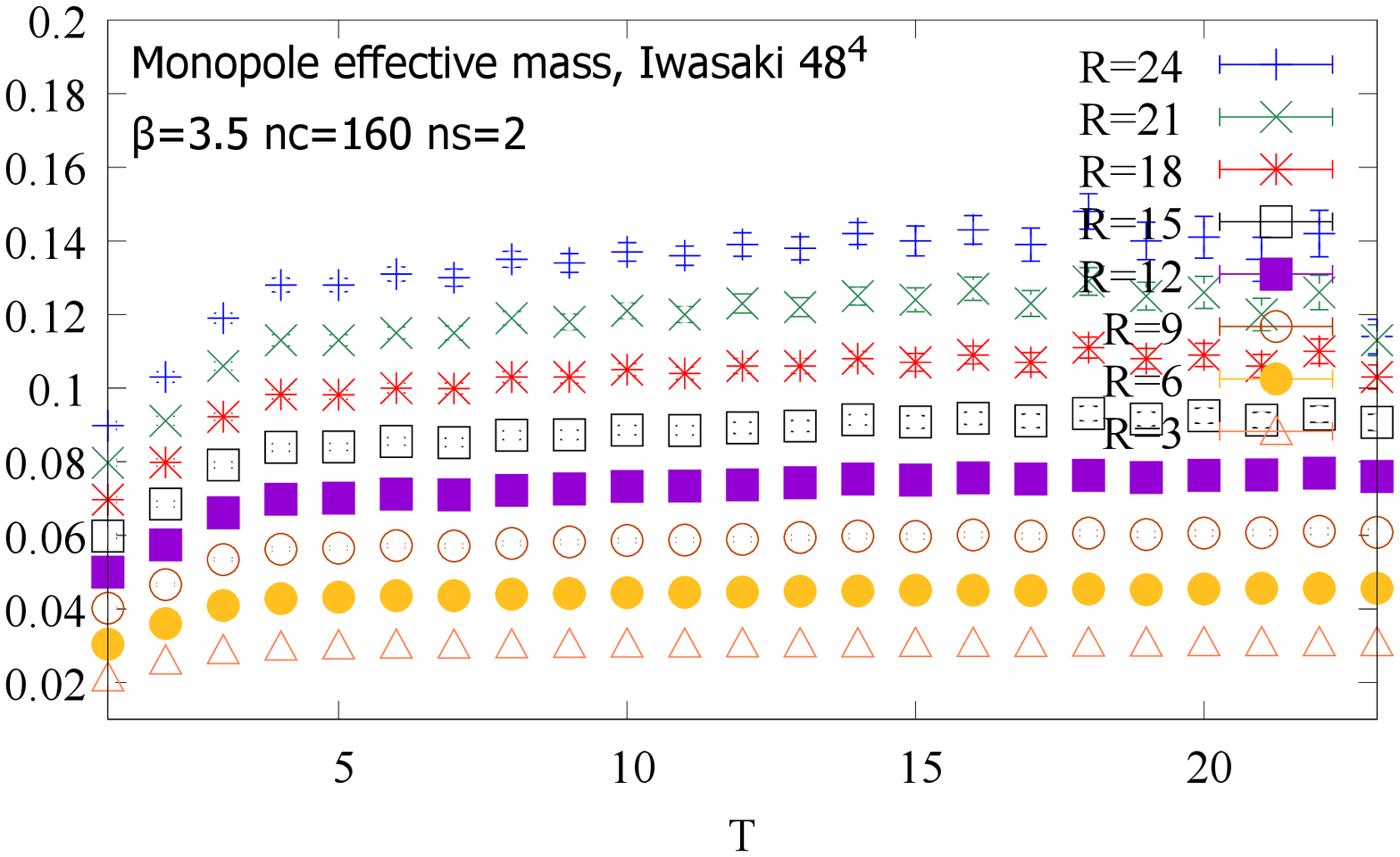}
  \end{minipage}
\end{figure}

\begin{figure}[htb]
\caption{Examples of the static potentials at $\beta=3.5$ on $48^4$ for  the Abelian (up) and the monopole (down) parts.}
\label{potential}
  \begin{minipage}[b]{0.9\linewidth}
    \centering
 \includegraphics[width=8cm,height=6.cm]{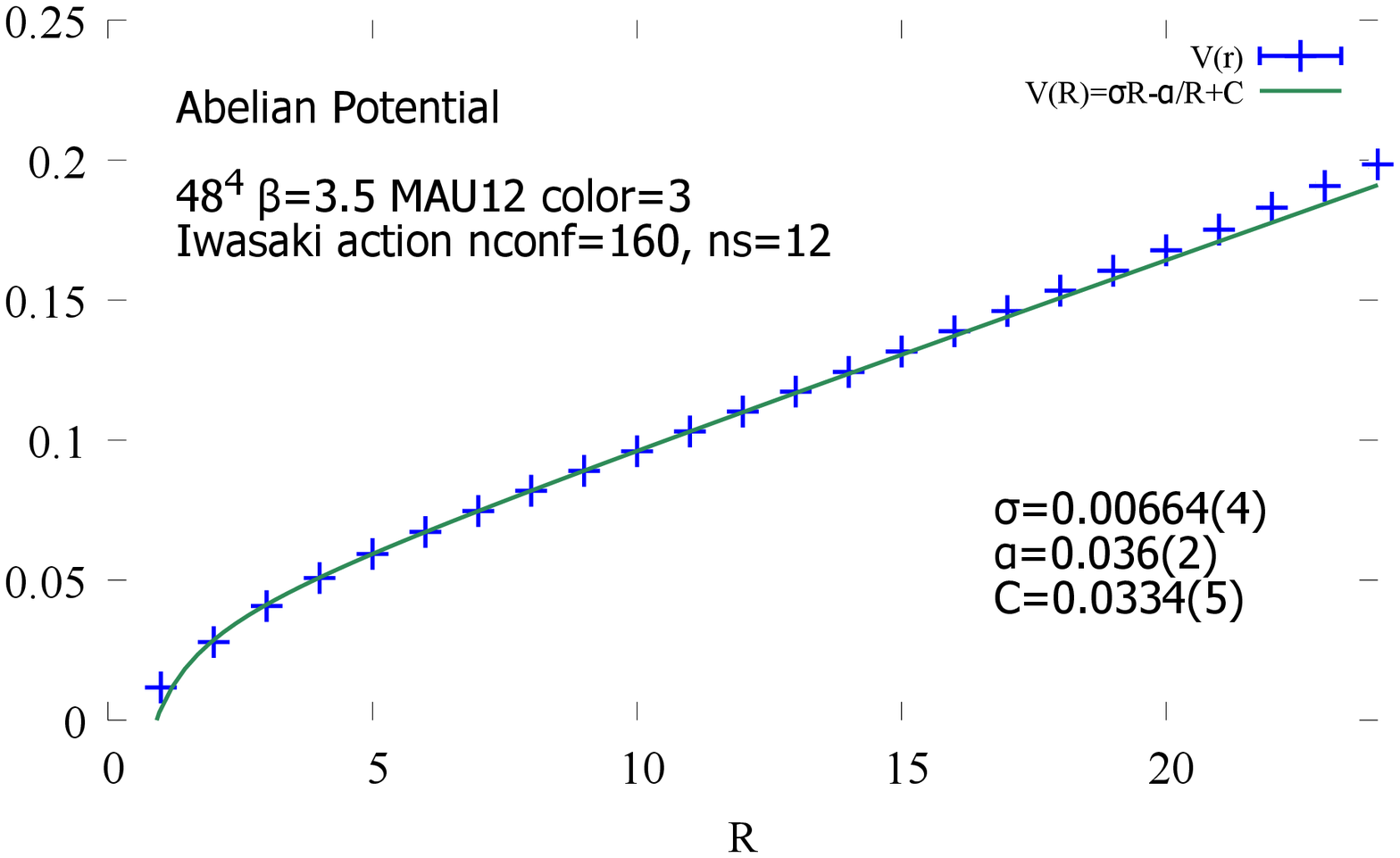}
  \end{minipage}
  \begin{minipage}[b]{0.9\linewidth}
    \centering
 \includegraphics[width=8cm,height=6.cm]{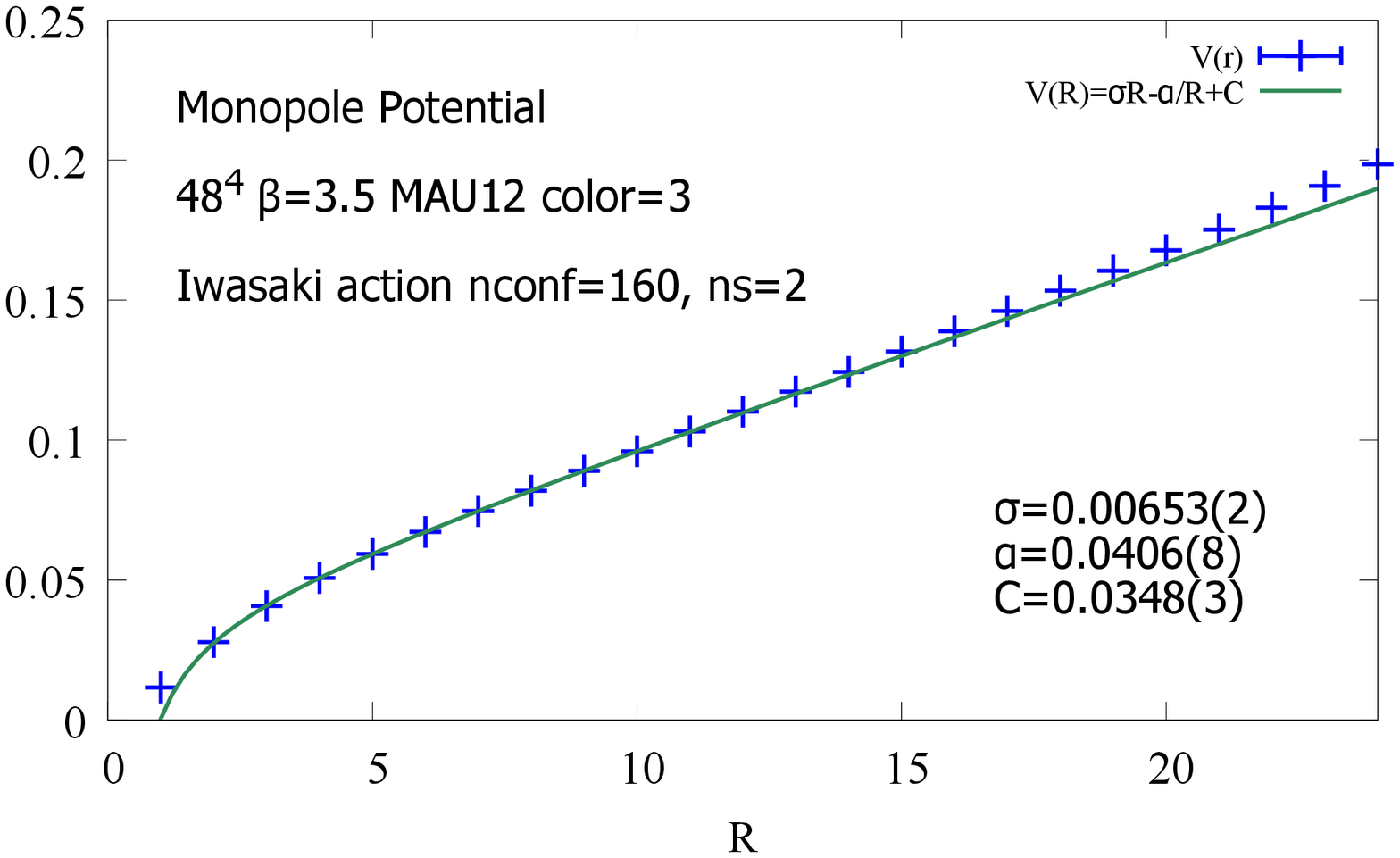}
  \end{minipage}
\end{figure}

\begin{figure}[htb]
\caption{Ratio of Abelian string tensions versus non-Abelian one on $48^4$.}
\label{sigAratio}  
    \centering
 \includegraphics[width=8cm,height=6.cm]{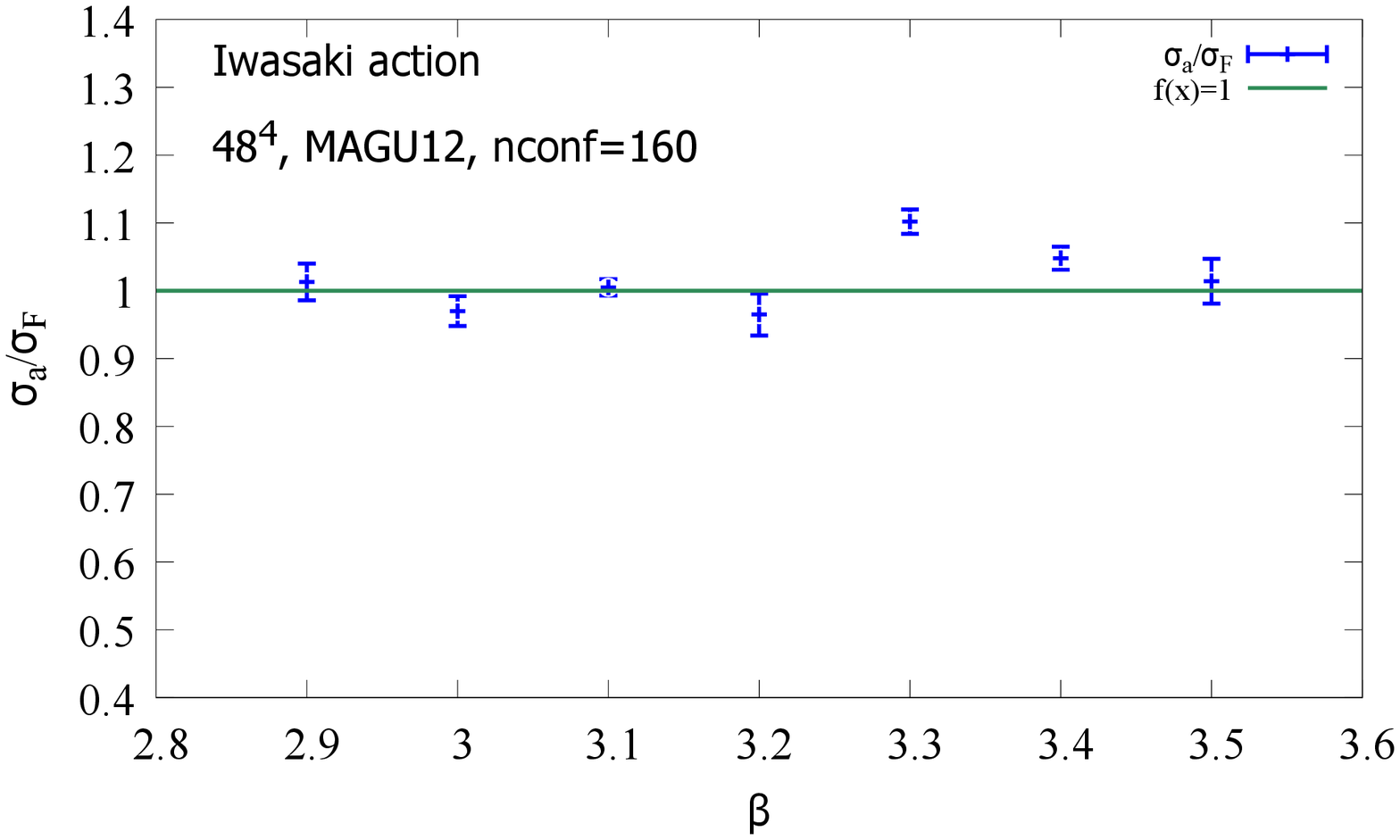}
\end{figure}


\begin{figure}[htb]
\caption{Ratio of monopole string tensions versus non-Abelian one on $48^4$.}
\label{sigMratio} 
    \centering
 \includegraphics[width=8cm,height=6.cm]{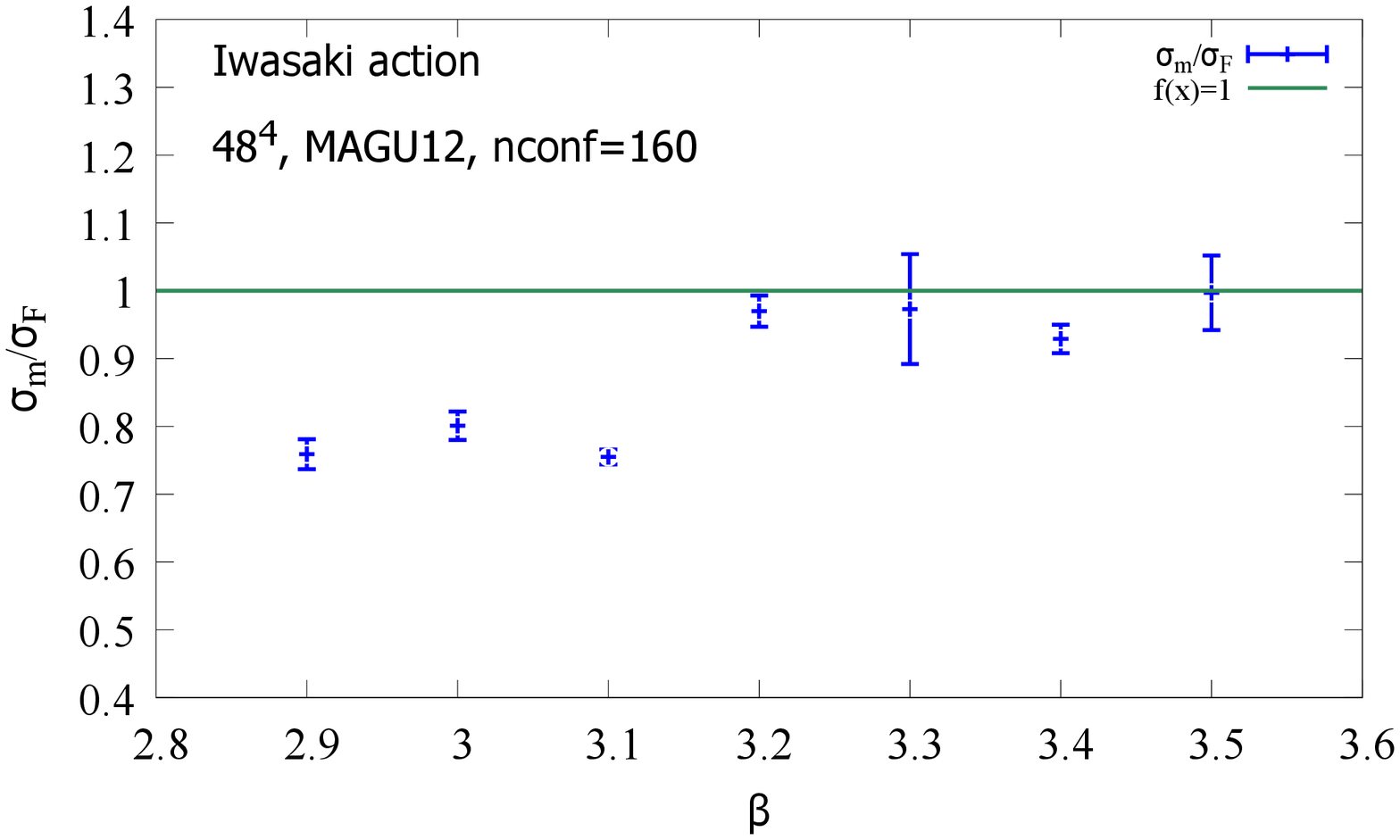}
\end{figure}

\section{Lattice settup of $SU(3)$ QCD} \label{lattice definition} 
To study the continuum limit clearly on large lattice volume, it is important to 
reduce the lattice artifact monopoles as much as possible and for that purpose,  
we adopt the maximally Abelian gauge (MA)~\cite{Kronfeld:1987ri,Kronfeld:1987vd,footnote} in which 
\begin{eqnarray}
R_{MA}(U(s,\mu))=\sum_s\sum_{\mu=1}^4 \tr(U^{\dag}(s,\mu)\vec{H}U(s,\mu)\vec{H}) \label{MAG}
\end{eqnarray}
is maximized under $SU(3)$ gauge transformations, where $\vec{H}$ is the diagonal Cartan subalgebra. 
After the MA gauge-fixing, we perform 
gauge-fixing with respect to the residual $U(1)^2$ symmetry in Landau gauge. Here we denote such serial gauge-fixings as MAU12.

Then Abelian link fields ${\theta^a_{\mu}(s)}$ and  Abelian Dirac-type monopoles on $SU(3)$ lattice are defined from non-Abelian link fields ${U_\mu(s)}$ as in the previous work~\cite{IHS:202207}. Maximizing the following quantity
\begin{eqnarray}
RA= \mathrm{Re} \tr\left\{\exp(i\theta_\mu^a(s)\lambda^a)U_\mu^{\dag}(s)\right\}, \label{RA}
\end{eqnarray}
where $\lambda^a$ is the Gell-Mann matrix leads us to, say, in the $\lambda^1$ case,  
\begin{eqnarray}
\theta^1_\mu(s)=\tan^{-1}\left\{\frac{\mathrm{Im}(U_{12}(s,\mu)+U_{21}(s,\mu))}{\mathrm{Re}(U_{11}(s,\mu)+U_{22}(s,\mu))}\right\}.  \label{Eq1}
\end{eqnarray}
 
To improve the overlapping, we perform the following smearings:\\
(1) The hypercubic smearing is done with respect to the temporal direction of non-Abelian link fields similarly as done in \cite{Hasenfratz:2001}. But the results are found to be not so sensitive on the hypercubic blocking.\\
(2) With respect to spaticial link variables, we perform Abelian smearing with the fixed smearing parameter $\alpha=2.3$ similarly as done with respect to non-Abelian link fields in Ref.\cite{Bali:1993}. We check the dependence of the iteration numbers of smearing $ns$ for $0\le ns\le 60$ on the behaviors of the effective mass and the overlap parameter. The results are not so different except for the small $t<3a$ or large $t>20a$. We show in Table \ref{potential_parameter} the simulation parameters.

We next define Abelian-like lattice monopoles. The unique reliable method ever known to define a lattice Abelian monopole is the one proposed in compact QED by DeGrand and Toussaint~\cite{DeGrand:1980eq} who utilize the fact that the Dirac monopole has a Dirac string with a magnetic flux satisfying the Dirac quantization condition~\cite{Dirac:1931}. Hence we adopt the method here, since the Abelian-like monopoles here are of the Dirac type in QCD. 

It is known that MA gauge fixing in $SU(3)$ has some ambiguities especially in defining Abelian monopoles correponding to the diagonal color components\cite{Stack:2001}. Here we adopt the simplest method in which two diagonal Gell-Mann matrices $\lambda_3$ and $\lambda_8$ are used. 

First we define Abelian plaquette variables  from the above Abelian link variables:
\begin{eqnarray}
\theta_{\mu\nu}^a(s)&\equiv&\partial_{\mu}\theta_{\nu}^a(s)-\partial_{\nu}\theta_{\mu}^a(s),
\label{abel_proj}
\end{eqnarray}
where $\partial_{\nu}(\partial'_{\nu})$ is a forward (backward) difference. Then the plaquette variable can be decomposed as follows:
\begin{eqnarray}
\theta_{\mu\nu}^a(s) &=&\bar{\theta}_{\mu\nu}^a(s)+2\pi
n_{\mu\nu}^a(s)\ \ (|\bar{\theta}_{\mu\nu}^a|<\pi),\label{abel+proj}
\end{eqnarray}
where $n_{\mu\nu}^a(s)$ is an integer
corresponding to the number of the Dirac string.
Then VNABI as Abelian-like monopoles is defined by
\begin{eqnarray}
k_{\mu}^a(s)&=& -\frac{1}{2}\epsilon_{\mu\alpha\beta\gamma}\partial_{\alpha}
\bar{\theta}_{\beta\gamma}^a(s+\hat\mu) \nonumber\\
&=&\frac{1}{2}\epsilon_{\mu\alpha\beta\gamma}\partial_{\alpha}
n_{\beta\gamma}^a(s+\hat\mu), \nonumber \\
J_{\mu}(s)&\equiv&\frac{1}{2}k_{\mu}^a(s)\lambda^a \label{eq:amon}.
\end{eqnarray}
This definition (\ref{eq:amon}) of VNABI satisfies the Abelian conservation condition and takes an integer value
which corresponds to the magnetic charge obeying the Dirac quantization
condition\cite{Dirac:1931}.  


\begin{table*}
\caption{\label{string_tension}
Simulation results of the Abelian and monopole string tensions $\sigma_{a,m}$ versus non-Abelian one $\sigma_F$. $ FR(r/a)$  is the fitting range. $V(r)=\sigma*r+c$ is used in the monopole fit at $\beta=2.9$ and $3.1$.
}
\begin{tabular}{c|c|c|c|c|c|c|c}
\hline
  &\multicolumn{3}{c|} {Abelian string tension}  & \multicolumn{3}{c|} {monopole string tension}& non-Abelian string tension\cite{Takeda:2004}   \\
$\beta$ & $\sigma_a$&$FR(r/a)$  &$\chi^2/N_{d.o.f}$& $\sigma_m$&$FR(r/a)$  &$\chi^2/N_{d.o.f}$& $\sigma_F$ \\
  \hline
2.9&0.02044(5) &(3,16)&0.12&0.01531(8)&(4,24)&1.16&0.02017(47)\\
3.0&0.01670(34)&(4,24)&0.79&0.01380(7)&(4,24)&0.96&0.01722(34)\\
3.1&0.01312(4)&(4,24)&1.67&0.00986(5)&(4,24)&1.24&0.01306(12)\\
3.2&0.01126(22)&(8,18)&1.38&0.01132(13)&(7,21)&0.96&0.01167(14)\\
3.3&0.00928(3)&(4,24)&0.89&0.00818(6)&(6,24)&0.874&0.00842(11)\\
3.4&0.007662(4)&(3,24)&0.06&0.00679(5)&(7,24)&0.93&0.00731(11)\\
3.5&0.00664(4)&(5,12)&1.01&0.00653(18)&(4,11)&0.72&0.00655(17)\\   
\end{tabular} 
\end{table*}

\section{Abelian and monopole static potentials}
We evaluate the static potentials from Abelian Wilson loops and their monopole contributions. Here, we take into account only a simple Abelian Wilson loop, say, of size $I \times J$. Then such an Abelian Wilson loop operator is expressed as
\begin{align}
W^a_{A}=\mathrm{exp}\{i\sum J^{ext,a}_{\mu}(s) \theta_{\mu}^a(s) \},  
\end{align}
where $J^{ext,a}_{\mu}(s)$ is an external electric current having a color $a$ taking $\pm1$ along the Wilson loop. Since $J^{ext,a}_{\mu}(s)$ is conserved, it is rewritten for such a simple Wilson loop in terms of an antisymmetric variable $M^a_{\mu\nu}$ as $J^{ext,a}_{\nu}=\partial'M^a_{\mu\nu}(s)$ with a forward (backward) difference $\partial_{\nu} (\partial'_{\nu})$. Note that $M^a_{\mu\nu}(s)$ take $\pm1$ on a surface with the Wilson loop boundary. Although we can choose any surface type, we adopt a minimal flat surface here. We get 
\begin{eqnarray}
 W^a_{A}=\mathrm{exp}\{-\frac{i}{2}\sum M^a_{\mu\nu}(s) \theta^{a}_{\mu\nu}(s)\}.
\end{eqnarray}
We investigate the monopole contribution to the static potential in order to examine the role of monopoles for confinement. The monopole part of the Abelian Wilson loop operator is extracted as follows \cite{Shiba:1994ab, SNW:1994}. Using the lattice Coulomb propagator $D(s-s')$, which satisfies $\partial_{\nu} \partial'_{\nu} D(s-s') = -\delta_{ss'}$, we get
\begin{eqnarray}
W^a_{A}&=& W^a_{mon} W^a_{ph} , \\
W^a_{mon}&=&\mathrm{exp} \{2 \pi i \sum k^a_{\beta}(s) \notag \\
&\times& D(s-s^{'}) \frac{1}{2} \epsilon_{\alpha \beta \rho \sigma } \partial_{\alpha} M^a_{\rho \sigma}(s^{'}) \} , \\
W^a_{ph}&=&\mathrm{exp}\{-i \sum \partial^{'}_{\mu} \bar{\theta}^a_{\mu \nu}(s) D(s-s^{'}) J^a_{\nu}(s^{'}) \} .
\end{eqnarray}
We then compute the static potential from the Abelian Wilson loops and the monopole Wilson loops in the MAU12 gauge on the $48^4$ lattices at $\beta =2.9, 3.0, 3.1, 3.2, 3.3, 3.4$ and $3.5$. They are shown as follows
\begin{eqnarray}
V^a_{A}(r)&=&-\textrm{lim}_{t\to\infty} \textrm{ln}<W^a_{A}>\\
V^a_{m}(r)&=&-\textrm{lim}_{t\to\infty} \textrm{ln}<W^a_{m}>.
\end{eqnarray}

We extract $V(r)$ from the least-squares fit with the single-
exponential form 
\begin{eqnarray}
W(r,t)=C(r)e^{-V(r)t}  \label{VR}
\end{eqnarray}
and choose the fit range of $t_1\le t\le  t_2$ such
that the stability of the so-called effective mass
\begin{eqnarray}
V^{eff}(r,t)=\textrm{ln}\frac{W(r,t)}{W(r,t+1)}
\end{eqnarray}
is observed in the range $t_1\le t\le t_2$\cite{SS:2014}. 
We also measure the overlap coefficient $C(r)$ in (\ref{VR}) to check if the ground-state part is extracted or not.
Then
we fit the potential to the usual functional form
  \begin{align}
    V_{fit}(r) = \sigma r -c/r + \mu , 
  \end{align}
where $\sigma$ denotes the string tension, $c$ the Coulombic coefficient, and $\mu$ the constant. 
Since the MA gauge breaks global color invariance, the Abelian and the monopole potentials depend on the color chosen. Here we show explicitly the color 3 diagonal components alone. 
Examples of the effective mass of the Abelian and the monopole Wilson loops are plotted in Fig.\ref{potential}. The fitting ranges as well as other lattice parameters in each case are described in Table~\ref{potential_parameter}.  
 
Examples of 
the Abelian and the monopole static potentials in the MAU12 at $\beta=3.5$ are shown in Fig.\ref{potential}.

 The results of the string tensions  on $48^4$ in the MAU12 gauge are summarized for various $\beta$ 
 in Table \ref{string_tension}
  and in Figs. \ref{sigAratio} and \ref{sigMratio}. Here  $\sigma_a$ and $\sigma_m$ are Abelian, and monopole string tensions which are compared with non-Abelian string tensions $\sigma_F$ determined in Ref\cite{Takeda:2004}.  Perfect Abelian dominance is seen quite well for all $2.9 \le \beta\le 3.5$ considered. The perfect Abelian dominance in MAU12 was shown also on $32^4$ lattices in the Wilson action\cite{SS:2014}. Table~\ref{string_tension} and Fig.\ref{sigAratio} show that the asymptotic scaling seems quite well satisfied. On the other hand, perfect monopole dominance seems satisfied  for $\beta \ge 3.2$ as seen from Table~\ref{string_tension} and Fig.\ref{sigMratio}. These results along with the previous results\cite{IHS:202207} without any additional gauge fixing but on smaller lattices are consistent with the expectation that the Abelian dual Meissner effect due to the new Abelian-like monopoles is the color confinement mechanism in the continuum limit. 
 
Some comments are in order. 
\begin{enumerate}
\item The errorbars of the Abelian and monopole potentials of color 3 in MAU12 are very small and so they are not clearly seen in Fig.\ref{potential}.
\item The errors in Table \ref{string_tension} are only statistical. There are systematic errors. Changes of the fitting range giving rise to one of the systematic errors are checked to be less than $10$ percent. 
  \item In the above, we show only the results with respect to the color 3 diagonal components. We also measure another color 8 diagonal components for all $\beta$. To extract the color 8 Abelian link fields from non-Abelian 
one needs to solve a quartic equation, so that a bit more complicated. But still we get 
almost good Abelian and monopole dominances. 
\item We also measure off-diagonal color components. But then the overlap coefficients $C(r)$ become smaller rapidly for large $r$ regions and then no Abelian and monopole dominances are observed. 
\item  The $SU(3)$ results in MAU12 obtained here seem better than those studied in the serial maximally Abelian and $U(1)$ Landau gauge (MAU1) and the maximally center gauge (MCG) in the $SU(2)$ case\cite{HIS:2020}. Note, however, in $SU(2)$, perfect Abelian and monopole dominances are clearly shown in the works without any additional gauge fixing\cite{Suzuki:2007jp,Suzuki:2009xy}. 
\end{enumerate}

\begin{table}[htb]
\begin{center}
\caption{\label{newconfigs}
Simulation parameters (Coupling constants $\beta$ of the Iwasaki action, 
the lattice size, the 
gauge configuration number used. The lattice spacing $a(\beta)$ are from Ref.\cite{Takeda:2004}.} 
\begin{tabular}{c|c|c|c|}
\hline
$\beta$ &volume& $N_{\rm conf}$& $a(\beta)$~[fm]\\ 
\hline
2.3 & $48^4$ & 80 & 0.1143(46)\\
2.4 & $48^4$ & 80 & 0.1143(46)\\
2.5 & $48^4$ & 80 & 0.1143(46)\\
2.6 & $48^4$ & 80 & 0.1080(60)\\
2.7 & $48^4$ & 80 & 0.0918(65)\\
2.8 & $48^4$ & 80 & 0.0855(75) \\
\hline
\end{tabular}
\end{center}
\end{table}

 \begin{figure}[htb]
\caption{Monopole density behaviors for fixed $n=1$ monopoles versus
$\beta$ (up) and for fixed $\beta$ versus $n$ (down).}
\label{beta-n-dependence}
  \begin{minipage}[b]{0.9\linewidth}
    \centering
 \includegraphics[width=8cm,height=6.cm]{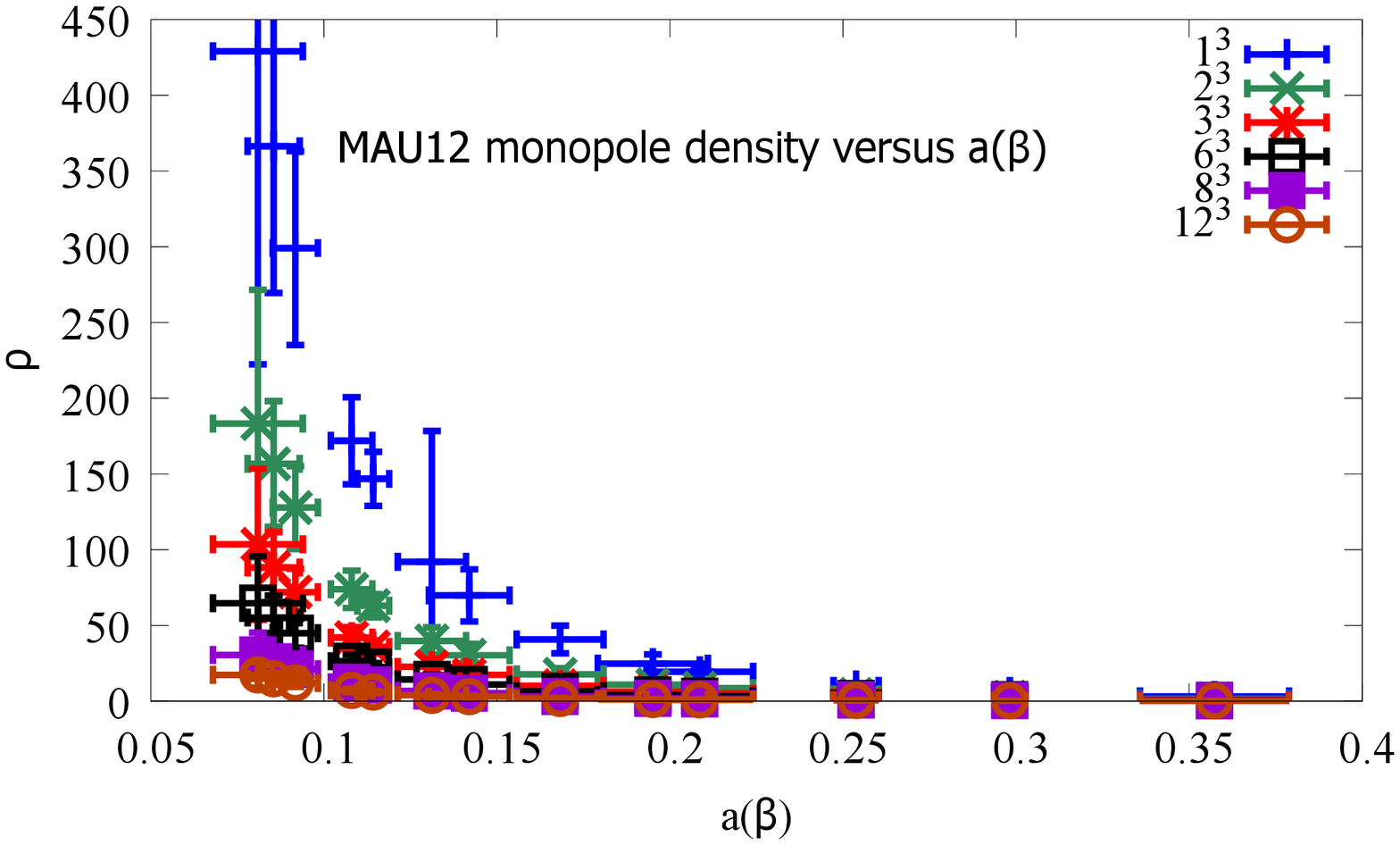}
  \end{minipage}
  \begin{minipage}[b]{0.9\linewidth}
    \centering
 \includegraphics[width=8cm,height=6.cm]{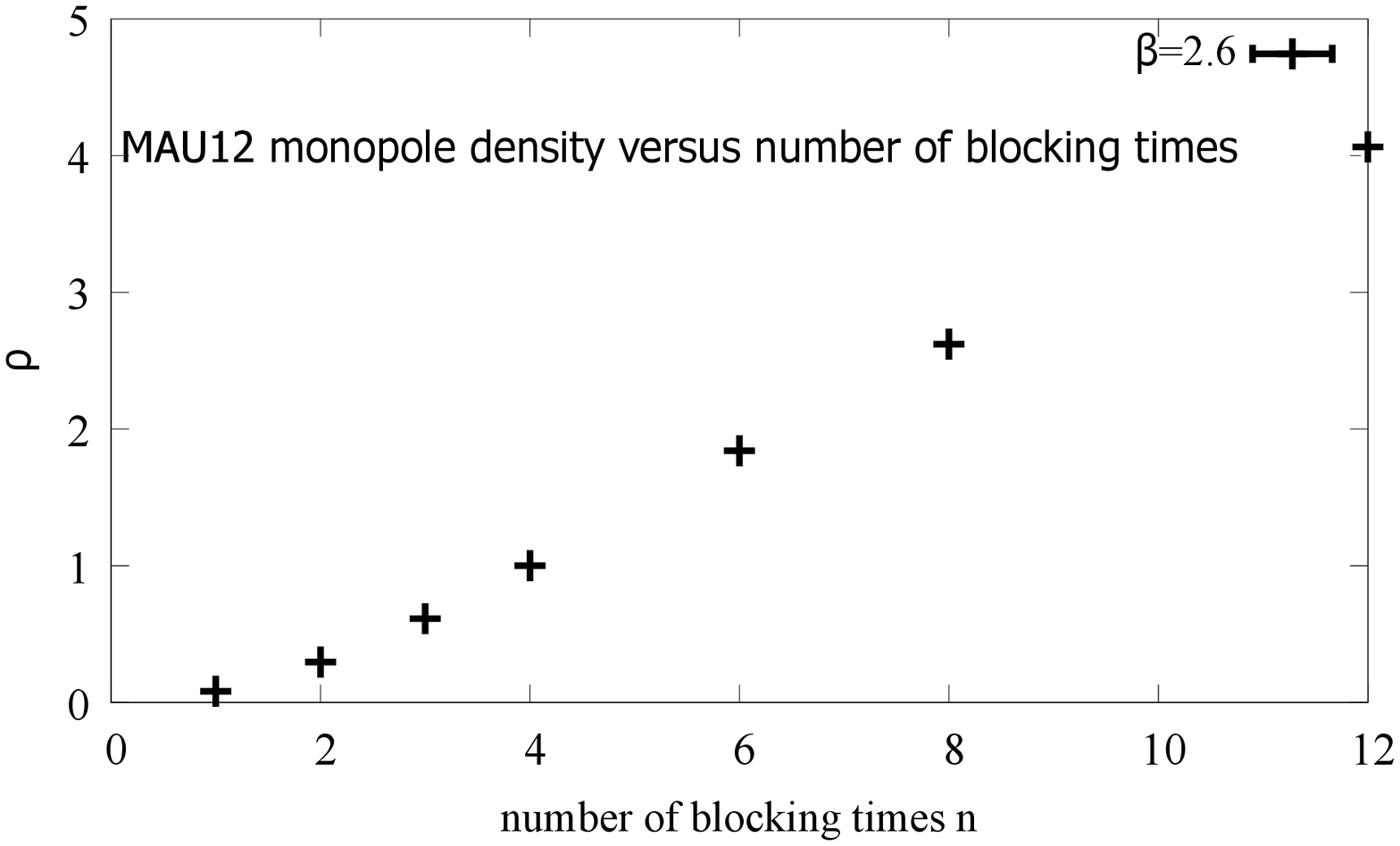}
  \end{minipage}
\end{figure}

\begin{figure}[htb]
\caption{Monopole density versus $b=na(\beta)$}
\label{b_dependence}
    \centering
 \includegraphics[width=8cm,height=6.cm]{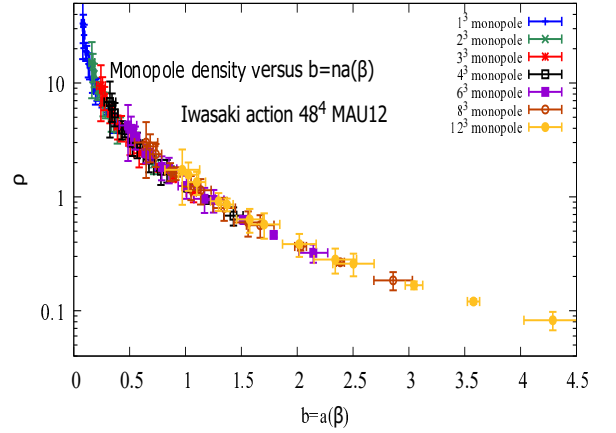}
\end{figure}

\section{Block spin transformation studies of the monopoles}
\subsection{The block spin transformation method}
Since Abelian monopoles considered here correspond to violation of non-Abelian Bianchi identity (VNABI) in the continuum\cite{Suzuki:2014wya,Suzuki:2017lco}, it is impossible to study the continuum limit of such quantities on lattice in the framework of the asymptotic scaling of usual continuum QCD where VNABI is assumed not to occur. Namely existence of line-like singulariries leading to VNABI is not assumed in the usual framework of QCD. To study the continuum limit of Abelian monopoles, therefore, one needs to adopt a completely different method. 
  
The renormalization-group method based on the block spin transformation is known to be a powerful tool for studying the continuum limit and critical phenomena especially in various spin-systems\cite{Kadanoff:1966,Jegerlehner:1976,Mack:1989}. When the original lattice has a volume $V$ with the lattice spacing $a$, the blocked lattice is defined as that having a lattice spacing $na$ on the lattice volume $V/n^3$ and the blocked spin is defined by integrating
out the original spins on the original lattice inside the blocked lattice. An infrared effective action is obtained describing the physics of the blocked spins leading us to the renormalization-group flow.     

The idea of the block spin with respect to Abelian monopoles on lattice was first introduced by Ivanenko et al.\cite{Ivanenko:1991wt} and applied to the study obtaining an infrared effective monopole action in Ref.\cite{Shiba:1994db}. The $n$ blocked monopole has a total magnetic charge inside the $n^3$ cube and is defined on
a blocked reduced lattice with the spacing $b=na$. 
%
The respective magnetic currents for each color are defined as
\begin{eqnarray}
k_{\mu}^{(n)}(s_n) &=& \frac{1}{2}\epsilon_{\mu\nu\rho\sigma}
\partial_{\nu}n_{\rho\sigma}^{(n)}(s_n+\hat{\mu}) \nonumber\\
    & = & \sum_{i,j,l=0}^{n-1}k_{\mu}(ns_n \nonumber\\
    &&  +(n-1)\hat{\mu}+i\hat{\nu}
     +j\hat{\rho}+l\hat{\sigma}), \label{excur}\\
n_{\rho\sigma}^{(n)}(s_n) &=& \sum_{i,j=0}^{n-1}
n_{\rho\sigma}(ns_n+i\hat{\rho}+j\hat{\sigma}),\nonumber
\end{eqnarray}
where $s_n$ is a site number on the reduced lattice and the color indices are not shown explicitly.
For example,
\begin{eqnarray*}
 k_{\mu}^{(2)}(s_2)&=&
\sum_{i,j,l=0}^{1}k_{\mu}(2s_2+\hat{\mu}+i\hat{\nu}
     +j\hat{\rho}+l\hat{\sigma}),\\
k_{\mu}^{(4)}(s_4)&=&\sum_{i,j,l=0}^{3}k_{\mu}(4s_4+3\hat{\mu}+i\hat{\nu}
     +j\hat{\rho}+l\hat{\sigma}) \nonumber \\
            &=&\sum_{i,j,l=0}^{1}k_{\mu}^{(2)}(2s_4+\hat{\mu}+i\hat{\nu}
     +j\hat{\rho}+l\hat{\sigma}).
\end{eqnarray*}
These equations show that the relation between $k_{\mu}^{(4)}(s_4)$ and $k_{\mu}^{(2)}(s_2)$ is similar to that between $k_{\mu}^{(2)}(s_2)$ and $k_{\mu}(s)$ and hence one can see the above equation (\ref{excur}) corresponds to the usual block spin transformation.
After the block spin transformation, the number of short lattice artifact  monopole loops decreases while loops having larger magnetic charges appear. For details, see Ref.\cite{Suzuki:2017lco}. \\

For the purpose of studying the scaling behaviors for wide range of $\beta$, we adopt the vacuum ensembles of the Iwasaki action  from $\beta=2.3$ till $\beta=2.8$ as shown in Table~\ref{newconfigs} in addition to those in Table~\ref{potential_parameter}.  For the additional range of $\beta$, we adopt only 80 configurations in Table~\ref{potential_parameter}, since the errors are very small
in the case of monopole density and the effective action studies.

\subsection{Monopole density}
The first observable is the gauge-invariant monopole density. If the 
Abelian monopoles exist in the continuum limit, the monopole density must exist non-vanishing in the continuum. In $SU(2)$, this seems to be realized actually~\cite{Suzuki:2017lco}.

In $SU(3)$ we have eight Abelian-like conserved monopole currents instead of three in $SU(2)$. Since monopoles are three-dimensional objects, the monopole density is defined as follows:
\begin{eqnarray}
\rho=\frac{\sum_{\mu,s_n}\sqrt{\sum_a(k_{\mu}^a(s_n))^2}}{4\sqrt{8}V_nb^3},\label{eq:Mdensity}
\end{eqnarray}
where $V_n=V/n^4$ is the 4 dimensional volume of the reduced lattice, $b=na(\beta)$ is the spacing of the reduced lattice after $n$-step block spin transformation.
The superscript $a$ denotes a color component. It is to be noted that we do not restrict ourselves only to the Abelian monopoles of color diagonal components as usually adopted in MAU12 gauge. Here we adopt Abelian monopoles of all color components and take the sum over all color components. Then  $\sum_a(k_{\mu}^a)^2$
is gauge-invariant in the continuum limit, since $J_\mu=k_\mu$ is an adjoint operator. Note that  we are studying the new Abelian-like monopoles of the Dirac type which must be independent of additional partial gauge fixing.\\

In general, the density $\rho$ is a function of two variables $\beta$ and $n$, i.e., $\rho=\rho(n,a(\beta))$.
When we change $\beta$ larger for fixed number of blocking step, the monopole density  decreases as shown in the upper part of Fig~\ref{beta-n-dependence} in the case of original unblocked monopole currents. No asymptotic scaling is seen for fixed number of blocking. On the otherhand, we change the number of blocking steps from $n=1$ to $n=12$, the monopole density increases monotonously for fixed $\beta$. \\
But it is interesting to show that, if we plot the monopole density versus  blocked lattice distance $b=na(\beta)$, we get a universal curve $\rho(n,a(\beta))\to \rho(b=na(\beta))$ depending on $b$ alone as shown in Fig.~\ref{b_dependence}.
There is a beautiful scaling function similarly as observed in $SU(2)$\cite{Suzuki:2017lco},
although the latter $SU(2)$ results have smaller errorbars and more appealing.
If the same behavior $\rho(b=na(\beta))$ is kept for $n\to \infty$, it correponds to the non-zero monopole density at $a(\beta)\to 0$, i.e., the continuum limit. Although we have studied the block spin transformation up to $n=12$, the results obtained support strongly existence of the continuum limit of the Abelian-like monopoles considered here
, since the asymptotic universal scaling function depending only on $b$ is realized.\\

In $SU(2)$, we have studied three other smooth gauge fixings as well as MAU1 and no gauge-dependence is
seen as expected from the new type of Abelian-like monopoles\cite{Suzuki:2014wya}. On the otherhand in $SU(3)$, we have not yet obtained another reliable gauge-fixed smooth vacuum ensemble except for those in MAU12. Hence to prove existence of the new type of Abelian-like monopoles in $SU(3)$, the scaling behavior in MAU12 alone is not enough. Gauge independence is still to be studied.

\begin{table}
\caption{The quadratic interactions used for the modified Swendsen method. Color index $a$ of the monopole current $k_{\mu}^a$ is omitted.}
\label{tbl:appquad} 
\begin{tabular}{cllcll}
{\it coupling $\mbra{F(i)}$} &  distance& $ \ \ \ \ \ \ \ \ $  {\it type} $ \ \ \ \ \ \ \ \ $  \\ 
\hline
$F(1)$    & (0,0,0,0) & $k_\mu(s)k_\mu(s)$  \\
$F(2)$    & (1,0,0,0) & $k_\mu(s)k_\mu(s+\hat{\mu})$  \\
$F(3)$    & (0,1,0,0) & $k_\mu(s)k_\mu(s+\hat{\nu})$ \\
$F(4)$    & (1,1,0,0) & $k_\mu(s)k_\mu(s+\hat{\mu}+\hat{\nu})$ \\
$F(5)$    & (0,1,1,0) & $k_\mu(s)k_\mu(s+\hat{\nu}+\hat{\rho})$ \\
$F(6)$    & (1,1,1,0) & $k_\mu(s)k_\mu(s+\hat{\mu}+\hat{\nu}+\hat{\rho})$ \\
$F(7)$    & (0,1,1,1) & $k_\mu(s)k_\mu(s+\hat{\nu}+\hat{\rho}+\hat{\sigma})$\\ 
$F(8)$    & (2,0,0,0) & $k_\mu(s)k_\mu(s+2\hat{\mu})$ \\
$F(9)$    & (1,1,1,1) & $k_\mu(s)k_\mu(s+\hat{\mu}+\hat{\nu}+\hat{\rho}+\hat{\sigma})$ \\
$F(10)$ & (0,2,0,0) & $k_\mu(s)k_\mu(s+2\hat{\nu})$ \\
\end{tabular}
\end{table}

\begin{figure}[htb]
\caption{The self-coupling constant $F(1)$ versus $b=na(\beta)$.}
\label{figF(1)}
    \centering
 \includegraphics[width=8cm,height=6.cm]
 {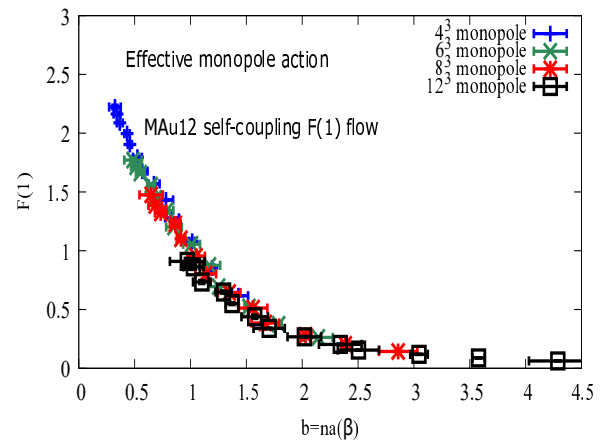}
\end{figure}
\begin{figure}[htb]
\caption{The nearest-neighbor coupling constant $F(2)$ versus $b=na(\beta)$.}
\label{figF(2)}
    \centering
 \includegraphics[width=8cm,height=6.cm]
 {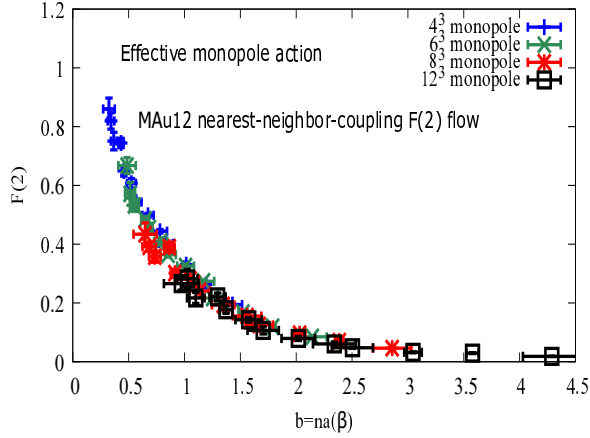}
\end{figure}
\begin{figure}[htb]
\caption{The another nearest-neighbor coupling constant $F(3)$ versus $b=na(\beta)$.}
\label{figF(3)}
    \centering
 \includegraphics[width=8cm,height=6.cm]
 {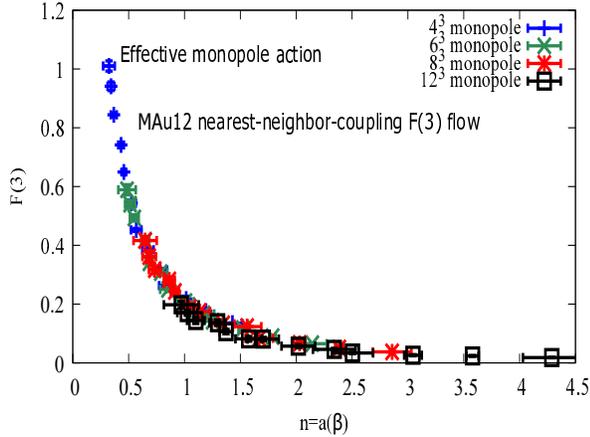}
\end{figure}
\begin{figure}[htb]
\caption{The next to nearest-neighbor coupling constant $F(4)$ versus $b=na(\beta)$.}
\label{figF(4)}
    \centering
 \includegraphics[width=8cm,height=6.cm]
 {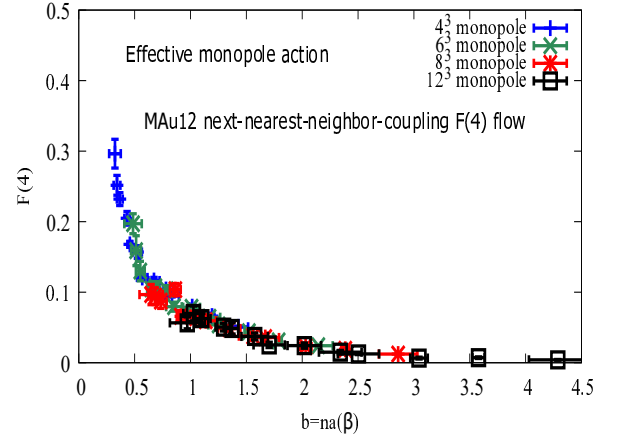}
\end{figure}

\subsection{Infrared effective monopole action}
The next observable is the infrared effective monopole action.
The effective action $S(k)$ for original monopoles $\{k_\mu^a(s)\}$  is defined  as follows:
\begin{eqnarray*}
e^{-{\cal S}[k]}&=&\int DU(s,\mu)e^{-S(U)}\\
&&\times \prod_a \delta(k_\mu^a(s)-\frac{1}{2}\epsilon_{\mu\nu\rho\sigma}\partial_\nu n_{\rho\sigma}^a(s+\hat{\mu})),
\end{eqnarray*}
where $S(U)$ is the Iwasaki gauge action.
The effective action for blocked monopoles $\{k_{\mu}^{(n)}(s_n)\}$ is evaluated as
\begin{eqnarray*}
&&e^{-{\cal S}[k^{(n)}]}=\Pi_{s,\mu}\sum_{k_\mu(s)=-\infty}^{\infty}
\delta(\partial'_{\mu}k_{\mu})e^{-{\cal S}[k]}\\
&&\times\delta(k_{\mu}^{(n)}(s_n)-\sum_{i,j,l=0}^{n-1}k_{\mu}(ns_n+(n-1)\hat{\mu}+i\hat{\nu}+j\hat{\rho}+l\hat{\sigma})).
\end{eqnarray*}
Then we get the renormalization flow of infrared effective monopole actions as ${\cal S}[k]\to {\cal S}[k^{(1)}]\to {\cal S}[k^{(2)}] \ldots $. 

Practically, we have to restrict the number of interaction terms of monopoles. 
It is natural to assume that monopoles which are far apart do not 
interact strongly and to consider only short-ranged local interactions 
of monopoles. 

We determine the monopole action~(\ref{eq:monoact}), that is, the set of couplings $F(i)$ from the monopole current 
ensemble $\mbra{k_{\mu}^a(s)}$ with the aid of an inverse 
Monte-Carlo method first developed by Swendsen~\cite{swendsen} 
and extended to closed monopole currents by Shiba and Suzuki 
~\cite{Shiba:1994db}. The details of the inverse Monte-Carlo method are reviewed in AppendixA of Ref.~\cite{Suzuki:2017zdh}. 

Also in $SU(3)$, we are dealing with Abelian-like monopoles of each color separately, the method is the same as done in $SU(2)$~\cite{Suzuki:2017zdh}. For simplicity, here we consider only the most important two-point interactions between monopole currents composed of first 10 couplings as infrared effective monopole action, since they are known as most important from the careful studies of $SU(2)$ case:
\begin{eqnarray}
 {\cal S}[k] = \sum_{i}^{10} F(i) {\cal S}_i[k], \label{eq:monoact}
\end{eqnarray}
where $F(i)$ are first 10 coupling constants shown explicitly in Table~\ref{tbl:appquad}.

Since we now consider vacuum configurations in the smooth MAU12 gauge, 
only the diagonal components are important. Hence, we consider only the monopole currents having a color 3.

As studied in the previous section discussing the monopole density, we perform the block spin transformation of monopole currents for $n=1,2,3,4,6,8,12$ on $48^4$ at $\beta=2.3\sim 3.5$ and try to fix the infrared monopole actions for all blocked monopoles. 
 
Contrary to the beautiful $SU(2)$ results~\cite{Suzuki:2017zdh},
the coupling constants for small steps of blocking $n=1\sim 3$ can not be determined well for  $\beta=2.3\sim 3.5$. We may need more delicate tuning of inital conditions for $F(i)$. Here we discuss only the results of the results of $F(i)$ for $n\ge 4$. All coupling constants are in general a function of $a(\beta)$ and $n$. But similarly as in the monopole density, the scaling behaviors are seen only when we plot $F(i)$ versus $b=na(\beta)$.  The most dominant self-coupling constant 
$F(1)$ is shown in Fig.\ref{figF(1)}. The result show that the coupling constant $F(1)$ is a function of $b=na(\beta)$ alone, namely the scaling behavior is seen. Behaviors of other important coupling constants are shown in Figs.\ref{figF(2)} $\sim$ \ref{figF(4)}. All data show similar scaling behaviors.

\section{Summary and discussions}
In this note, the scaling behaviors of the new Abelian-like monopoles in pure $SU(3)$ QCD are studied adopting the Iwasaki improved gauge action for wide range of $\beta$ and the number of blocking transformations from $n=1,2,3,4,6,8,12$. To reduce lattice-artifact monopoles, we adopt here the maximally Abelian gauge and $U1^2$ Landau gauge. 
\begin{enumerate}
  \item The perfect Abelian dominance and the perfect monopole dominance are seen fairly well with respect to Abelian and monopole string tensions. The 
asymptotic
scaling behaviors are observed roughly in these cases. The $SU(3)$ results here look better than those in $SU(2)$\cite{HIS:2020}.
  \item The block spin transformation studies with respect to Abelian monopoles are done. The behaviors of the monopole densities $\rho(n,a(\beta))$ of the blocked monopole currents show the beautiful scaling behavior: $\rho(n,a(\beta)) \to \rho(b=na(\beta))$, i.e. $\rho$ is a function of $b=na(\beta)$ alone. The scaling behaviors are seen here for $n=1,2,3,4,6,8,12$. If on larger lattices, similar scaling behaviors are seen for $n\to\infty$, it means $a(\beta)\to 0$, the continuum limit. It is stressed that, although we adopt MAU12 gauge, the scaling behavior of the monopole density is seen with respect to $SU(3)$ invariant combination summing over all color components. 
  \item Adopting the inverse Monte Carlo method, we determine the coupling constant flow of the effective monopole action under the blocking transformation.
Although we restrict ourselves to important two-point monopole current interactions, we get the scaling behaviors also. Namely, all coupling constants which usually a two-point function of $n$ and $a(\beta)$ are actually found to be a function of $b=a(\beta)$ alone.
\item It is interesting to know what is the continuum theory of Abelian monopoles. The present author along with colleagues has studied the continuum theory of Abelian monopoles. An Abelian dual Higgs model\cite{Suzuki:1988} seems to be the theory of Abelian monopoles in the continuum limit.  See the references\cite{KKIS:2003,CIS:2004}.  
\item
These results are all on $48^4$ lattice for various coupling constants of the Iwasaki gauge action, adopting MAU12 gauge for reducing the lattice-artifact monopoles. It is absolutely necessary to show gauge independence to prove the new type of Abelian monopoles coming from the violation of non-Abelian Bianchi identity at least as done in $SU(2)$\cite{Suzuki:2007jp,Suzuki:2009xy} without adopting any additional gauge fixing. But such studies in $SU(3)$  seem at present almost impracticable except for the previous study on a small lattice\cite{IHS:202207}. Hence it is desirable to study in  smooth gauges other than MAU12 as done in $SU(2)$ case\cite{Suzuki:2017lco,Suzuki:2017zdh}. We have tried the Maximal Center (MCG) gauge\cite{DelDebbio:1996mh,DelDebbio:1998uu}, since in $SU(2)$ it shows after the simulated annealing\cite{Bornyakov:2000ig} a similar scaling behavior as in MAU1. But in $SU(3)$ at present the simple MCG gauge fixing is too difficult to find the real maximum point. There seem to exist so many local maxima in the MCG gauge funtional. Such a work will be done in future.
\end{enumerate}

\section*{Acknowledgements}
This work used High Performance Computing resources provided by Cybermedia Center of Osaka University through the JHPCN System Research Project (Project ID: jh220002). The numerical simulations of this work were done also using High Performance Computing resources at Research Center for Nuclear Physics  of Osaka University, at Cybermedia Center of Osaka University. The author would like to thank these centers for their support of computer facilities. This work is finacially supported by JSPS KAKENHI Grant Number JP19K03848.


\begin{thebibliography}{99}
\bibitem{tHooft:1975pu}
G.~'t~Hooft,
\newblock in {\em Proceedings of the EPS International}, edited by A.~Zichichi,
  p. 1225, 1976. 
\bibitem{Mandelstam:1974pi}
S.~Mandelstam,
\newblock Phys. Rept. {\bf 23}, 245 (1976).
\bibitem{tHooft:1981ht}
G.~'t~Hooft,
\newblock Nucl. Phys. {\bf B190}, 455 (1981).
\bibitem{Kronfeld:1987ri}
A.~S. Kronfeld, M.~L. Laursen, G.~Schierholz, and U.~J. Wiese,
\newblock Phys. Lett. {\bf B198}, 516 (1987).
\bibitem{Kronfeld:1987vd}
A.~S. Kronfeld, G.~Schierholz, and U.~J. Wiese,
\newblock Nucl. Phys. {\bf B293}, 461 (1987).
\bibitem{footnote}
The reader may wonder why  some smearing  or cooling methods smoothing the vacuum 
are 
 not used 
instead of introducing a partial gauge-fixing. But these methods are not useful in reducing lattice artifact monopoles but keeping physical monopoles unchanged. 
Only smooth non-local gauge-fixngs like MAG seem to be able to do the work. 
\bibitem{Suzuki:1992rw}
T.~Suzuki,
\newblock Nucl. Phys. Proc. Suppl. {\bf 30}, 176 (1993).

\bibitem{Chernodub:1997ay}
M.~N. Chernodub and M.~I. Polikarpov,
\newblock 
\newblock in {\em "Confinement, Duality and Nonperturbative Aspects of QCD"},
  edited by P.~van Baal, p. 387, Cambridge, 1997, Plenum Press.
\bibitem{Shiba:1994ab}
H. ~Shiba and T. ~Suzuki,
\newblock Phys. Lett. B{\bf 333}, 461 (1994).
\bibitem{SNW:1994}
J. D. Stack, S. D. Neiman, and R. J. Wensley, Phys. Rev.
{\bf D 50}, 3399 (1994).


\bibitem{Bonati:2010tz}
C.~Bonati, A.~Di~Giacomo, L.~Lepori and F.~Pucci, 
\newblock Phys. Rev. {\bf D81}, 085022 (2010).
\bibitem{Suzuki:2014wya}
T. Suzuki, A new scheme for color confinement due to violation of the non-Abelian Bianchi identities,  hep-lat: arXiv:1402.1294 (2014)
\bibitem{Suzuki:20220422}
T.Suzuki, 
\newblock Monopoles of the Dirac type and color confinement in QCD - $SU(3)$ invariant picture of color confinement -, arXiv:2204.11514.
\bibitem{Arafune:1974uy}
J.~Arafune,  P.G.O.~Freund  and C.J.~Goebel, 
\newblock J.Math.Phys. {\bf 16}, 433 (1975).


J.~Arafune,  P.G.O.~Freund  and C.J.~Goebel, 
\newblock J.Math.Phys. {\bf 16}, 433 (1975).

\bibitem{Dirac:1931} 
P. Dirac,  Proc. Roy. Soc. (London) A 133, 60 (1931).



\bibitem{Suzuki:2007jp}
T.~Suzuki, K.~Ishiguro, Y.~Koma and T.~Sekido,
\newblock Phys. Rev. {\bf D77}, 034502 (2008).

\bibitem{Suzuki:2009xy}
T.~Suzuki, M.~Hasegawa, K.~Ishiguro, Y.~Koma and T.~Sekido,
\newblock Phys. Rev. {\bf D80}, 054504 (2009).

\bibitem{Suzuki:2017lco}
T.~Suzuki, K.~Ishiguro and V.~Bornyakov, 
Phys. Rev. {\bf D97}, 034501 (2018); Phys. Rev. {\bf D97}, 099905(E) (2018).

\bibitem{Suzuki:2017zdh}
T.~Suzuki, 
Phys. Rev {\bf D97}, 034509 (2018).

\bibitem{IHS:202207}
K.Ishiguro, A.Hiraguchi and T. Suzuki,Phys Rev. {\bf D106} .014515 (2022), arXiv:2207.04436.
\bibitem{Luscher:2001up}
M.~L\"{u}scher and P.~Weisz, JHEP {\bf 0109}, 010 (2001), hep-lat/0108014.

\bibitem{Luscher2002}
M.~L\"{u}scher and P.~Weisz, JHEP {\bf 0207}, 049 (2002), hep-lat/0207003.

\bibitem{Elitzur:1975}
S.~Elitzur,
tat\newblock Phys. Rev. {\bf D12}, 3978 (1975).


\bibitem{Iwasaki:1985}
Y. Iwasaki, Nucl. Phys. B258 (1985) 141; Univ. of Tsukuba report
UTHEP-118 (1983), unpublished.

\bibitem{IKKY:1997}
Y. Iwasaki, K. Kanaya, T. Kaneko and T. Yoshie,
\newblock Phys. Rev. {\bf D56}, 151 (1997).	arXiv:hep-lat/9610023.

\bibitem{Takeda:2004}
S.~Takeda et al. (CP-PACS Collaboration), Phys. Rev. D 70, 074510 (2004).
\bibitem{Koma:2003}
Y. Koma, M. Koma, E.-M. Ilgenfritz, T. Suzuki and M.I. Polikarpov,
Phys.Rev. D68 (2003) 094018, arXiv: hep-lat/0302006.
\bibitem{Stack:2001}
J. D. Stack, W. W. Tucker, R. J. Wensley, 
\newblock The Maximal Abelian Gauge, Monopoles, and Vortices in $SU(3)$ Lattice Gauge Theory, 	arXiv:hep-lat/0110196, 2001.
\bibitem{Hasenfratz:2001}
A.~Hasenfratz and F.~Knecht, Phys.\ Rev.\ D {\bf 64}, 034504 (2001).  [arXiv:hep-lat/0103029].

\bibitem{Bali:1993}
G. S. Bali and K. Schilling, Phys. Rev. {\bf D 47}, 661 (1993).

\bibitem{DeGrand:1980eq}
T.~A. DeGrand and D.~Toussaint,
\newblock Phys. Rev. {\bf D22}, 2478 (1980).
\bibitem{SS:2014}
N.~Sakumichi and H.~Suganuma, Phys. Rev. {\bf 90}, 111501 (2014).

\bibitem{HIS:2020}
A.Hiraguchi,K.Ishiguro and T.Suzuki, 
Phys. Rev. {\bf D102} 114504 (2020),arXiv:2011.14377

\bibitem{Kadanoff:1966}
L.P.~Kadanoff, Physics {\bf 2} 2, 1966.
\bibitem{Jegerlehner:1976}
F.~Jegerlehner, An Introduction to the Theory of Critical Phenomena and the Renoramlization Group, Lecture in "Trois\`{e}me cycle de la physique en Suisse Romande" (the Ecole Polytechnique 
F\`{e}d\`{e}rale de Lausanne, May 1976) 
\bibitem{Mack:1989}
G.~Mack, Multigrid Methods in Quantum Field Theory, Cargese Lecturs 1987, in Nonpertubative Quantum Field Theory (Plenum, NY, 1989)

\bibitem{Ivanenko:1991wt}
T.L.~Ivanenko, A. V.~Pochinsky and M.I.~Polikarpov,
\newblock  Phys. Lett. {\bf B302}, 458 (1993). 
\bibitem{Shiba:1994db}
H.~Shiba and T.~Suzuki,
\newblock Phys. Lett. {\bf B351}, 519 (1995).

\bibitem{swendsen} 
R.H. Swendsen,Phys. Rev. Lett. 
{\bf 52},1165 (1984).
\bibitem{Suzuki:1988}
T.~Suzuki, Prog. Theor. Phys. {\bf 80}, 929 (1988). S.~Maedan and T.~Suzuki, Prog. Theor. Phys. {\bf 81}, 229 (1989). 

\bibitem{KKIS:2003}
Y.~Koma, M.~Koma, E.-M.~Ilgenfritz and T.~Suzuki, Phys. Rev. {\bf D68},
114504 (2003).
\bibitem{CIS:2004}
M.N.~Chernodub, K.~Ishiguro and T.~Suzuki, Phys. Rev. {\bf D69}, 094508 (2004).

\bibitem{DelDebbio:1996mh}
L. ~Del Debbio, M. ~Faber, J. ~Greensite and S. ~Olejnik,
\newblock	Phys. Rev. \textbf{D55}, 2298 (1997)
\bibitem{DelDebbio:1998uu}
L. ~Del Debbio, M. ~Faber, J. ~Giedt, J. ~Greensite and S. ~Olejnik,
\newblock	Phys. Rev. \textbf{D58}, 094501 (1998)
\bibitem{Bornyakov:2000ig}
V. G. ~Bornyakov, D. A. ~Komarov and M.I.~Polikarpov,
\newblock Phys. Lett. {\bf B497}, 151 (2001).



\end{thebibliography}
\end{document}